\documentclass[%
 reprint,
superscriptaddress,
showpacs,showkeys,preprintnumbers,
 amsmath,amssymb,
 aps,
]{revtex4-1}

\usepackage{graphicx}
\usepackage{dcolumn}
\usepackage{bm}
\usepackage{here}
\usepackage[dvips]{color}
\usepackage{multirow}

\newcommand{\abs}[1]{\left\vert{#1}\right\vert}

\makeatletter
  \def\erf{\mathop{\operator@font erf}\nolimits}
  \def\erfc{\mathop{\operator@font erfc}\nolimits}
  \def\Erf{\mathop{\operator@font Erf}\nolimits}
  \def\Shi{\mathop{\operator@font Shi}\nolimits}
  \def\Chi{\mathop{\operator@font Chi}\nolimits}
  \def\Ei{\mathop{\operator@font Ei}\nolimits}
  \def\cosec{\mathop{\operator@font cosec}\nolimits}
  \def\sech{\mathop{\operator@font sech}\nolimits}
  \def\cosech{\mathop{\operator@font cosech}\nolimits}
  \newcommand\hypgeo[2]{{}_{#1}{\operator@font F}_{#2}}
  \def\Re{\mathop{\operator@font Re}\nolimits}
  \def\Im{\mathop{\operator@font Im}\nolimits}
\makeatother

\begin{document}


\title{Energy dependent angular distribution of individual $\bm{\gamma}$-rays in the $^{139}$La($\bm{n}$, $\bm{\gamma}$)$^{140}$La* reaction}

\def\affNagoya{Nagoya University, Furocho, Chikusa, Nagoya 464-8602, Japan}
\def\affKyushu{Kyushu University, 744 Motooka, Nishi, Fukuoka 819-0395, Japan}
\def\affJAEA{Japan Atomic Energy Agency, 2-1 Shirane, Tokai 319-1195, Japan}
\def\affTokyoTech{Tokyo Institute of Technology, Meguro, Tokyo 152-8551, Japan}
\def\affRCNP{Osaka University, Ibaraki, Osaka 567-0047, Japan}

\author{T.~Okudaira}
\affiliation{\affNagoya}

\author{S.~Endo}
\affiliation{\affNagoya}
\affiliation{\affJAEA}

\author{H.~Fujioka}
\affiliation{\affTokyoTech}

\author{K.~Hirota}
\affiliation{\affNagoya}

\author{K.~Ishizaki}
\affiliation{\affNagoya}

\author{A.~Kimura}
\affiliation{\affJAEA}

\author{M.~Kitaguchi}
\affiliation{\affNagoya}

\author{J.~Koga}
\affiliation{\affKyushu}

\author{Y.~Niinomi}
\affiliation{\affNagoya}

\author{K.~Sakai}
\affiliation{\affJAEA}

\author{T.~Shima}
\affiliation{\affRCNP}

\author{H.~M.~Shimizu}
\affiliation{\affNagoya}

\author{S.~Takada}
\affiliation{\affKyushu}

\author{Y.~Tani}
\affiliation{\affTokyoTech}

\author{T.~Yamamoto}
\affiliation{\affNagoya}

\author{H.~Yoshikawa}
\affiliation{\affRCNP}

\author{T.~Yoshioka}
\affiliation{\affKyushu}


\date{\today}


\begin{abstract}
Neutron energy-dependent angular distributions were observed for individual $\gamma$-rays from the 0.74~eV p-wave resonance of $^{139}$La+$n$ to several lower excited states of $^{140}$La.
The $\gamma$-ray signals were analyzed in a two dimensional histogram of the $\gamma$-ray energy, measured with distributed germanium detectors, and neutron energy, determined with the time-of-flight of pulsed neutrons, to identify the neutron energy dependence of the angular distribution for each individual $\gamma$-rays.
The angular distribution was also found for a photopeak accompanied with a faint p-wave resonance component in the neutron energy spectrum. Our results can be interpreted as interference between s- and p-wave amplitudes which may be used to study discrete symmetries of fundamental interactions. 

\end{abstract}

\pacs{13.75.Cs, 
21.10.Hw,
21.10.Jx,
21.10.Re,
23.20.En,
24.30.Gd,
24.80.+y,
25.40.Fq,
25.70.Gh,
27.60.+j,
29.30.Kv
}
\keywords{compound nuclei,
partial wave interference,
neutron radiative capture reaction}
\maketitle


\section{Introduction}
In the neutron absorption reaction of $^{139}$La, extremely large parity violation with a size of (9.56 $\pm$ 0.35)\% was observed in the 0.74~eV p-wave resonance located on a tail of an s-wave resonance~\cite{alf83}. It is understood that fundamental parity violation in the nucleon-nucleon interaction of the order of 10$^{-7}$~\cite{pot74,yua86,ade85} is enhanced by the interference between s- and p-wave amplitudes, referred to as  s-p mixing model~\cite{sus82,sus82E}.
The interference introduces a neutron energy-dependent angular distribution of $\gamma$-rays in the vicinity of a p-wave resonance with respect to the incident neutron momentum~\cite{fla85}.
The neutron energy dependence of the angular distribution for individual $\gamma$-rays was previously  measured using a germanium detector assembly and the intense neutron beam at the Japan Proton Accelerator Research Complex (J-PARC)~\cite{Okudaira18}. A clear angular distribution depending on the incident neutron energy was found in the vicinity of the p-wave resonance for the $\gamma$-rays resulting from the transition of the p-wave resonance of $^{139}\rm{La}+${\it n} to the ground state of $^{140}$La. \par
In this paper, the angular distributions of the $\gamma$-rays resulting from the p-wave resonance to lower excited states of $^{140}$La are studied as a function of the incident neutron energy by applying the same analysis method as in Ref.~\cite{Okudaira18} to photopeaks for transitions to the lower excited states. 
\\\\\\\\\\\\\\
\section{Experiment}
\subsection{Experimental Setup}
The data set used in this paper is the same as that used to evaluate the angular distribution of $\gamma$-rays to the ground state. The measurement of the angular distribution of the individual $\gamma$-rays in the $^{139}$La($n$,$\gamma$)$^{140}$La reaction was conducted using an intense pulsed neutron beam and a germanium detector assembly at beamline 04 of the Materials and Life Science Experimental Facility (MLF) at J-PARC~\cite{ANNRI}.  The germanium detector assembly consists of 22 high quality germanium crystals, pointing at angles from 36$^\circ$ to 144$^\circ$ with respect to the incident neutron beam direction~\cite{kim12}. A natural-abundant lanthanum metal plate with dimensions of 1~mm $\times$ 40~mm $\times$ 40~mm was placed at the detector center. The distance from the moderator surface to the La target is 21.5~m. The proton beam power was 150~kW and measurement time was 60 hours. A more detailed description of the experiment is given in Ref.~\cite{Okudaira18}.

\subsection{Measurement}
\begin{figure}[htbp]
	\centering
		\includegraphics[width=0.9\linewidth]{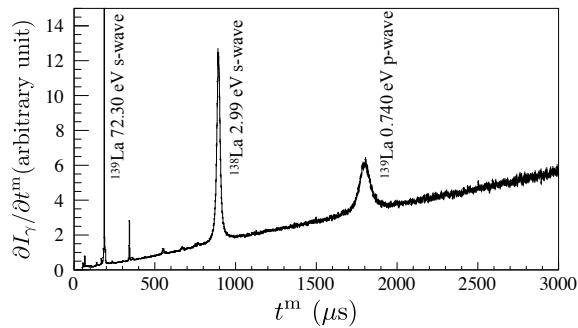}
	\caption[]{
	Neutron TOF spectrum defined as $\partial I_{\gamma}/\partial t^{\rm m}$. It is normalized relative to the incident beam intensity as a function of $t^{\rm m}$.
	}
	\label{TOF}
\end{figure}

\begin{figure}[htb]
	\centering
	\includegraphics[width=0.9\linewidth]{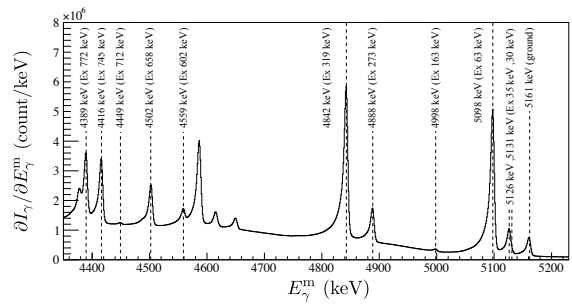}
	\caption[]{
	Expanded $\gamma$-ray spectrum defined as $\partial I_{\gamma}/\partial E_\gamma^{\rm m}$. The dotted line shows the literature value of the photopeak energy. Three photopeaks around 4600~keV are single escape peaks from the $\gamma$-rays of 5161 keV, 5131 keV, 5126 keV, and 5098~keV. }
	\label{Gamma}
\end{figure}

\begin{figure}[htbp]
	\centering
	\includegraphics[width=0.7\linewidth]{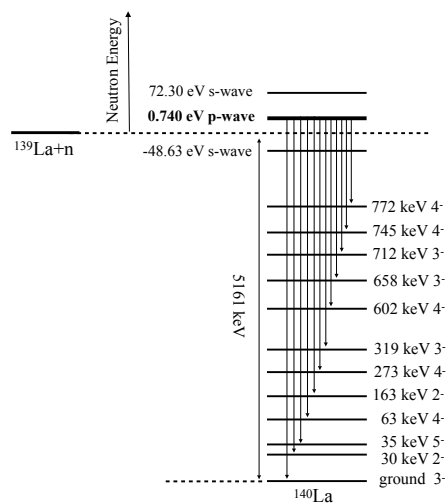}
	\caption[]{
	Transitions from $^{139}$La+$n$ to $^{140}$La. The dashed line shows separation energy of $^{139}$La+$n$. The transitions can actually occur not only from the p-wave resonance, but also from the s-wave resonances.  
	}
	\label{TransitionLa}
\end{figure}

The same variables defined in Ref.~\cite{Okudaira18} are used to evaluate the angular distributions of the $\gamma$-rays in this paper.  
The deposit energy of the $\gamma$-rays in the germanium crystal $E_{\gamma}^{\rm m}$ is obtained from the pulse height. The detection time of the $\gamma$-rays $t^{\rm m}$ is measured from the timing pulse of the injection of the proton beam bunch. These are obtained for each $\gamma$-ray event. The variable $t^{\rm m}$ corresponds to the Time Of Flight (TOF) of the incident neutrons for prompt $\gamma$-rays, and the corresponding neutron energy $E_{n}^{\rm m}$ is calculated using $t^{\rm m}$. The neutron energy in the center-of-mass system $E_{n}$ is defined as well. The total number of $\gamma$-ray events detected in the experiment are denoted as $I_{\gamma}$. 
A 2-dimensional histogram corresponding to $\partial^2I_{\gamma}/\partial t^{\rm m}\partial E^{\rm m}_\gamma$ was obtained for each germanium crystal as the experimental result.
The histogram of  $\partial I_{\gamma}/\partial t^{\rm m}$, which corresponds to a neutron TOF spectrum, is shown in Fig.~\ref{TOF}. The $\gamma$-ray events are integrated for $E^{\rm m}_\gamma \ge 2\,{\rm MeV}$ to remove delayed $\gamma$-rays. It is normalized relative to the incident beam spectrum for $t^{\rm m}$, which is obtained from a measurement of the 477.6~keV $\gamma$-rays in the neutron absorption reaction of ${}^{10}$B with an enriched ${}^{10}$B target.
The small peak at $t^{\rm m}\sim1800~\mu\rm{s}$ is the p-wave resonance, and the $1/v$ component is mainly derived from the tail of an s-wave resonance in the negative energy region as listed in Table~\ref{tab:res}.
\begin{table}[bp]
\begin{center}
	\begin{tabular}{c||c|c|c|c|c|}
	$r$ &
	$E_r\,[{\rm eV}]$ & 
	$J_r$ & 
	$l_r$ & 
	$\Gamma^{\gamma}_r\,[{\rm meV}]$ &
	$g_{ r}\Gamma^{\rm{n}}_r\,[{\rm meV}]$ 
	\\
	\hline
	$1$ & $-48.63^{\rm (a)}$ & $4^{\rm (a)}$ & $0$ &$62.2^{\rm (a)}$ &$(571.8)^{\rm (a) \ast}$ \\
	$2$ & $0.740\pm 0.002^{\rm (b)}$ & $4^{\rm (b)}$ & $1$ & $40.41\pm 0.76^{\rm (b)}$ & $(5.6 \pm0.5 )\times 10^{-5}$$^{\rm (c)}$ \\
	$3$ & $72.30\pm0.05^{\rm (c)}$ & $3^{\rm (b)}$ & $0$ & $75.64\pm2.21^{\rm (d)}$ & $11.76\pm0.53^{\rm (d)}$ \\
	\end{tabular}
	\caption{
	Resonance parameters of the neutron resonances of $^{139}$La+$n$. The resonance parameters $E_r$, $J_r$, $l_r$, $\Gamma_r^\gamma$, $g_r$ ,and $\Gamma_r^n$ are resonance energy, total angular momentum, orbital angular momentum, $\gamma$ width, {\it{g}}-factor and neutron width, respectively. Parameter $r$ denotes the resonance number. (a) taken from Ref.~\cite{mughabghab} and Ref.~\cite{JENDLall}.
	(b) taken from Ref.~\cite{Okudaira18}. (c) taken from Ref.~\cite{JENDL-La-28}. 
	(d) calculated from Refs.~\cite{JENDL-La-26} and \cite{JENDL-La-28}.
	$^{\ast}$The neutron width for the negative resonance was calculated using $\abs{E_1}$ instead of $E_1$.
	}
	\label{tab:res}
\end{center}	
\end{table}
Figure~\ref{Gamma} shows a $\gamma$-ray spectrum defined as $\partial I_{\gamma}/\partial E_{\gamma}^{\rm m}$. Photopeaks of $\gamma$-ray transitions of $^{139}{\rm{La}}(n,\gamma)$$^{140}\rm La$ reactions to the lower excited and ground states are observed in Fig.~\ref{Gamma}. A schematic diagram of the level scheme of the $^{139}$La($n$,$\gamma$)$^{140}$La reaction is also shown in Fig.~\ref{TransitionLa}~\cite{NNDC}.

\begin{figure*}[htbp]
	\begin{center}
	\includegraphics[width=180mm]{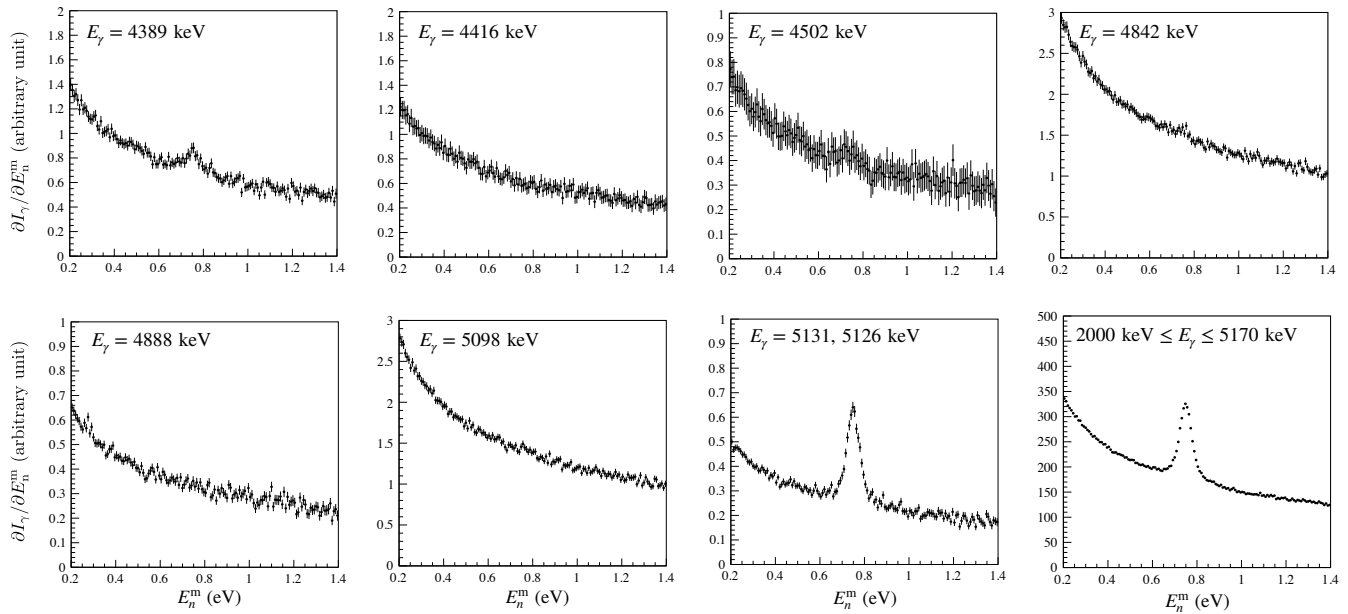}
	\caption[]{Neutron spectra gated with each photopeak and inclusive $\gamma$-rays. The energy of photopeaks or the gated region of $\gamma$-ray energy show at the top left of each histogram. }
	\label{n_spectra}
	\end{center}
\end{figure*}

Here, we focus on the intense transitions to the lower excited states of 30~keV, 35~keV, 63~keV, 273~keV, 319~keV, 658~keV, 745~keV, 772~keV, and the inclusive $\gamma$-ray transitions.  Histograms of $\partial I_{\gamma}/\partial E^{\rm m}_{n}$, which correspond to neutron energy spectra, gated with each photopeak and inclusive $\gamma$-rays are shown in Fig.~\ref{n_spectra}. We can see that the 0.74~eV p-wave resonance appears in several transitions. Note that the spectra are the sum of all detector angles. The photopeak region was taken as the full width at the quarter maximum.  Since the photopeaks of the 5126~keV and 5131~keV completely overlap, they are considered to be one photopeak. As a gated energy region for inclusive $\gamma$-rays, $2000~{\rm{keV}} \leq E^{\rm m}_{\gamma} \leq 5170~\rm{keV}$ was taken. 
The spectra were normalized by the incident neutron beam spectrum measured with the boron target. The background caused by the Compton-scattered $\gamma$-rays for each photopeak was estimated by a third-order polynomial fit in the low and high energy region of each photopeak for each detector and subtracted. Since a loss of 2\% of the total $\gamma$-ray counts occurred due to the DAQ system, a loss correction was also applied~\cite{Okudaira18}.

\subsection{Angular Distribution}\label{Ang}
\begin{figure*}[htbp]
	\begin{center}
	\includegraphics[width=180mm]{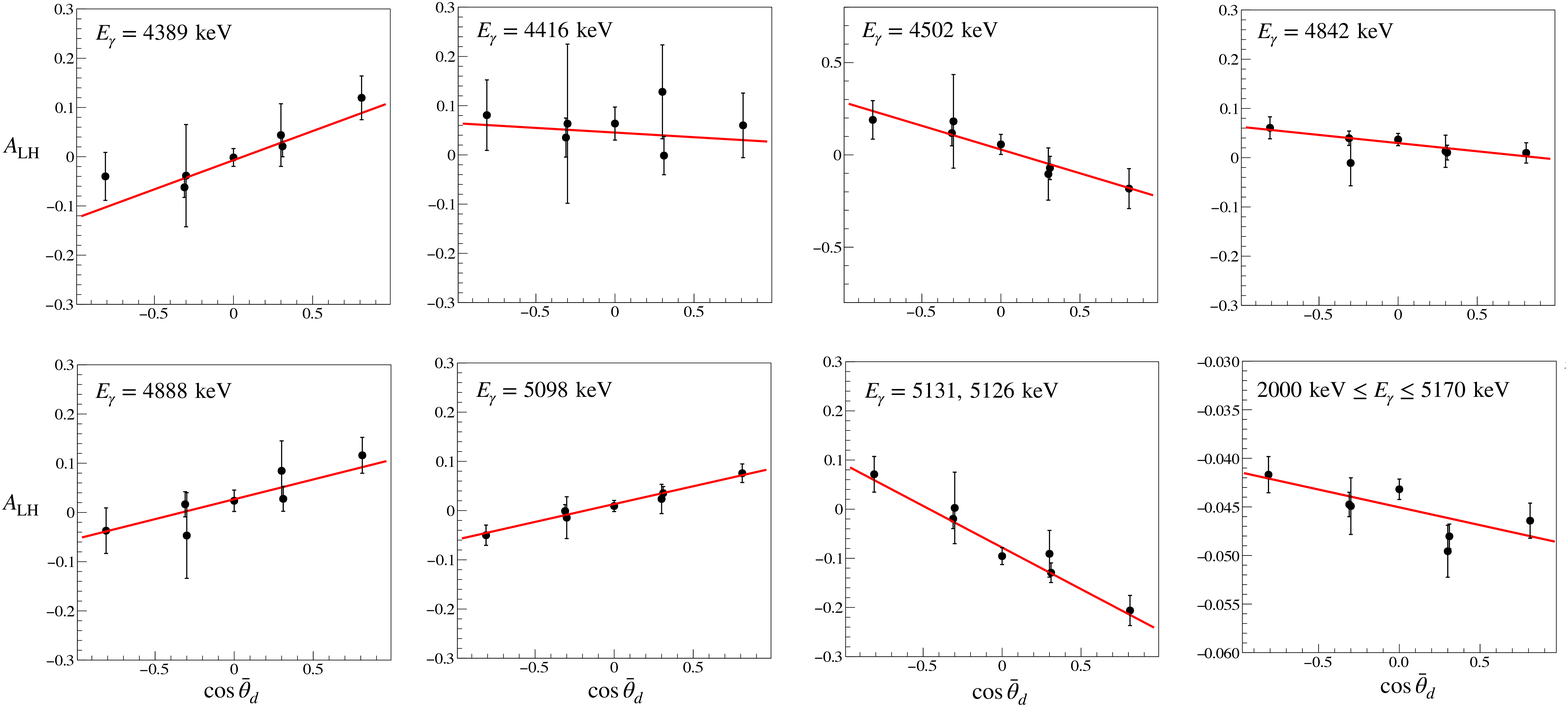}
	\caption[]{Angular distributions of $A_{\rm{LH}}$ for each photopeak and inclusive $\gamma$-rays. The solid lines are the fit results.}
	\label{ALH}
	\end{center}
\end{figure*}

The neutron-energy dependence of the angular distribution causes an asymmetric resonance shape in the neutron energy spectrum, which is measured by the asymmetry parameter defined as
\begin{eqnarray}
A_{\rm{LH}}(\theta_d)=\frac{N_{\rm L}(\theta_d)-N_{\rm H}(\theta_d)}{N_{\rm L}(\theta_d)+N_{\rm H}(\theta_d)},
\end{eqnarray}
where $\theta_d$ is the detector angle with respect to the incident neutron momentum, and $N_{\rm L}$ and $N_{\rm H}$ are integrals in the region of  $E_{\rm 2}-2\Gamma_{\rm 2} \le E_{n} \le E_{\rm 2}$ and $E_{\rm 2}\le E_{n} \le E_{\rm 2}+2\Gamma_{\rm 2}$, respectively. Variables $E_{\rm 2}$ and $\Gamma_{\rm 2}$ denote the resonance energy and total width of the p-wave resonance, which is defined by the $\gamma$ width and neutron width shown in Table~\ref{tab:res} as $\Gamma_{\rm 2}=\Gamma^\gamma_2+\Gamma^n_2$.  The asymmetry is plotted for effective detector angle $\bar{\theta}_d$, which is obtained with a simulation of the germanium detector assembly~\cite{Takada18}, and fitted using a function of the form $A_{\rm {LH}}(\bar{\theta}_{d})=A\cos\bar{\theta}_d+B$ with free parameters $A$ and $B$. The angular distributions of $A_{\rm{LH}}$ for the photopeaks and inclusive $\gamma$-rays are shown in Fig.~\ref{ALH}. 
The fit results of $A$, which correspond to the angular distribution of the asymmetry, are listed with the photopeak energy $E_\gamma$, excitation energy $E_{\rm{ex}}$, and the angular momentum of the final state $F$ in the Table~\ref{tab:ang}. 
Non-zero angular distributions were found in the transition to the excited state of 30~keV and/or 35~keV, 63~keV, 658~keV, and 772~keV with a confidence level of over 99.7\% . 

\begin{table}[htbp]
\begin{center}
	\begin{tabular}{c|c|c|c|}	
	$E_{\gamma}$ [keV] &
	$E_{\rm ex}$ [keV]& 
	$F$ & 
	$A$ 
	\\
	\hline
	4389 &772 & $4,5,6 $ & $0.118 \pm 0.030$ \\
	4416 & 745& $4 $ & $-0.020 \pm 0.049$ \\
	4502 &658 & $3 $ & $-0.257 \pm 0.078$ \\
	4842 &319 & $3 $ & $-0.033 \pm 0.016$ \\
	4888 &273 & $4 $ & $0.081 \pm 0.030$ \\
	5098 &63 & $4 $ & $0.072 \pm 0.015$ \\
	$5131, 5126$ & 30, 35 & $2,5$ & $-0.169\pm0.020$ \\
	$5161$ & 0 & $3$ & $-0.388\pm0.024$~\cite{Okudaira18} \\
	inclusive & - & - & $-0.0037\pm0.0014$\\
	\end{tabular}
	\caption{
	The fit results of $A$ for each photopeak and inclusive $\gamma$-rays. 
	}
	\label{tab:ang}
\end{center}	
\end{table}

\section{Discussion}
\subsection{Photopeak at 5098~keV}
\begin{figure}[htbp]
	\centering
	\includegraphics[width=1.0\linewidth]{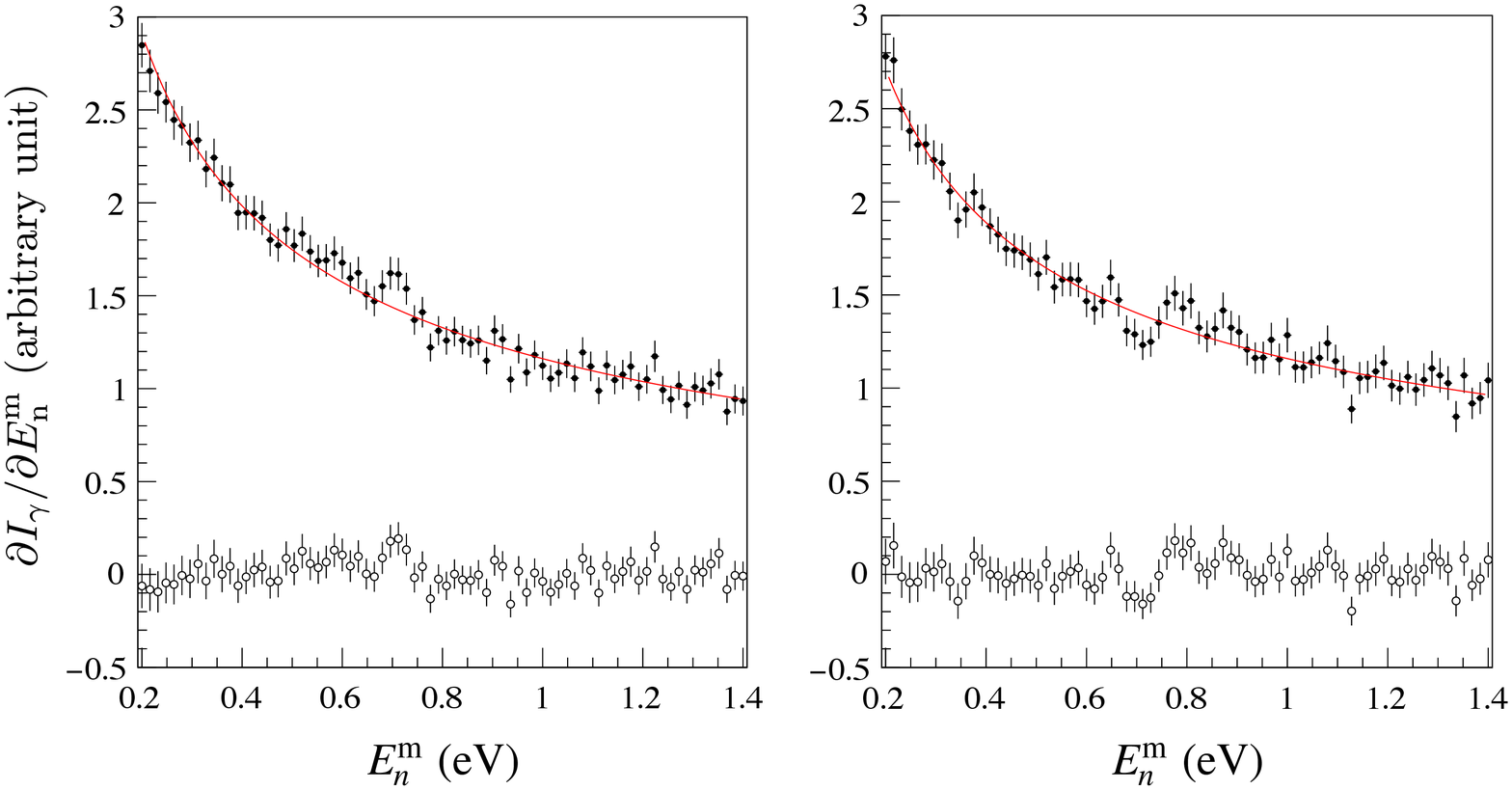}
	\caption[]{Neutron energy spectra gated with the 5098~keV photopeak for 36$^{\circ}$ (left) and 144$^{\circ}$ detectors (right). Solid lines show fit results to the s-wave component. Black points and white points show the spectra before and after subtraction of the s-wave component, respectively.}
	\label{36_144deg}
\end{figure}

Although the p-wave resonance does not appear in the neutron energy spectrum gated with 5098~keV photopeak as shown in Fig.~\ref{n_spectra}, the angular distribution $A$ is observed with a confidence level of over 99.7\%. This phenomenon can also be confirmed using the neutron energy spectra for 36$^\circ$ and 144$^\circ$ detectors shown in Fig.~\ref{36_144deg}. 
In Fig.~6, the s-wave component in the spectra, which obeys the $1/v$ law, is fitted using $f(E_n^{\rm{m}})=a/\sqrt{E_n^{\rm{m}}}+b$ with free parameters $a$ and $b$ for the regions except for the p-wave resonance, and the neutron energy spectra before and after the subtraction of the s-wave component are shown. We can see that they have a slight asymmetric shape at 0.74~eV, and moreover, the shapes reverse with respect to 0.74~eV for 36$^{\circ}$ and 144$^{\circ}$ detectors. This angular-dependent asymmetric component has also been observed for 5161~keV photopeak as shown in Fig.~13 in Ref.~\cite{Okudaira18} and is attributed to the interference term between s- and p-wave amplitudes. The significant value of $A$ observed for faint $\gamma$-ray transition from the p-wave resonance can be understood as the result of the moderately weak transition amplitude of $\gamma$-rays via the p-wave component; the magnitude of $A$ is proportional to the product of the transition amplitudes of s- and p-wave components while the $\gamma$-ray intensity is proportional to the square of that of p-wave component. This interpretation is clarified by the differential cross section of ($n$,$\gamma$) reactions based on the s-p mixing model described in APPENDIX E in Ref.~\cite{Okudaira18}.  The ordinary p-wave resonance shown in Fig.~\ref{n_spectra} corresponds to the second term of $a_0$ in Eq.~(E7) in Ref.~\cite{Okudaira18}, which has no angular distribution. In contrast, $a_1$, which is the interference term between s- and p-wave amplitudes, produces the angular-dependent asymmetric shape at the p-wave resonance for the detector angle satisfying $\cos \theta_\gamma \ne 0$ as shown in  Eq.~(E1) and Eq.~(E7) in Ref.~\cite{Okudaira18}. The $a_1$ term in Eq.~(E7) in Ref.~\cite{Okudaira18} is proportional to the branching ratio from the p-wave resonance to the $f$-th $\gamma$-ray transition $\lambda_{2f}$, whereas the second term of $a_0$ is proportional to $\lambda_{2f}^2$. Consequently, the second term of $a_0$ can be suppressed compared to $a_1$ when the branching ration from the p-wave resonance to the particular final state is small. In this way, the phenomenon that the 5098~keV photopeak has the angular distribution while no p-wave resonance appears can be explained by the interference term $a_1$ and the suppression of the $a_0$ amplitude due to the branching ratio.

\begin{figure*}[htbp]
	\begin{center}
	\includegraphics[width=180mm]{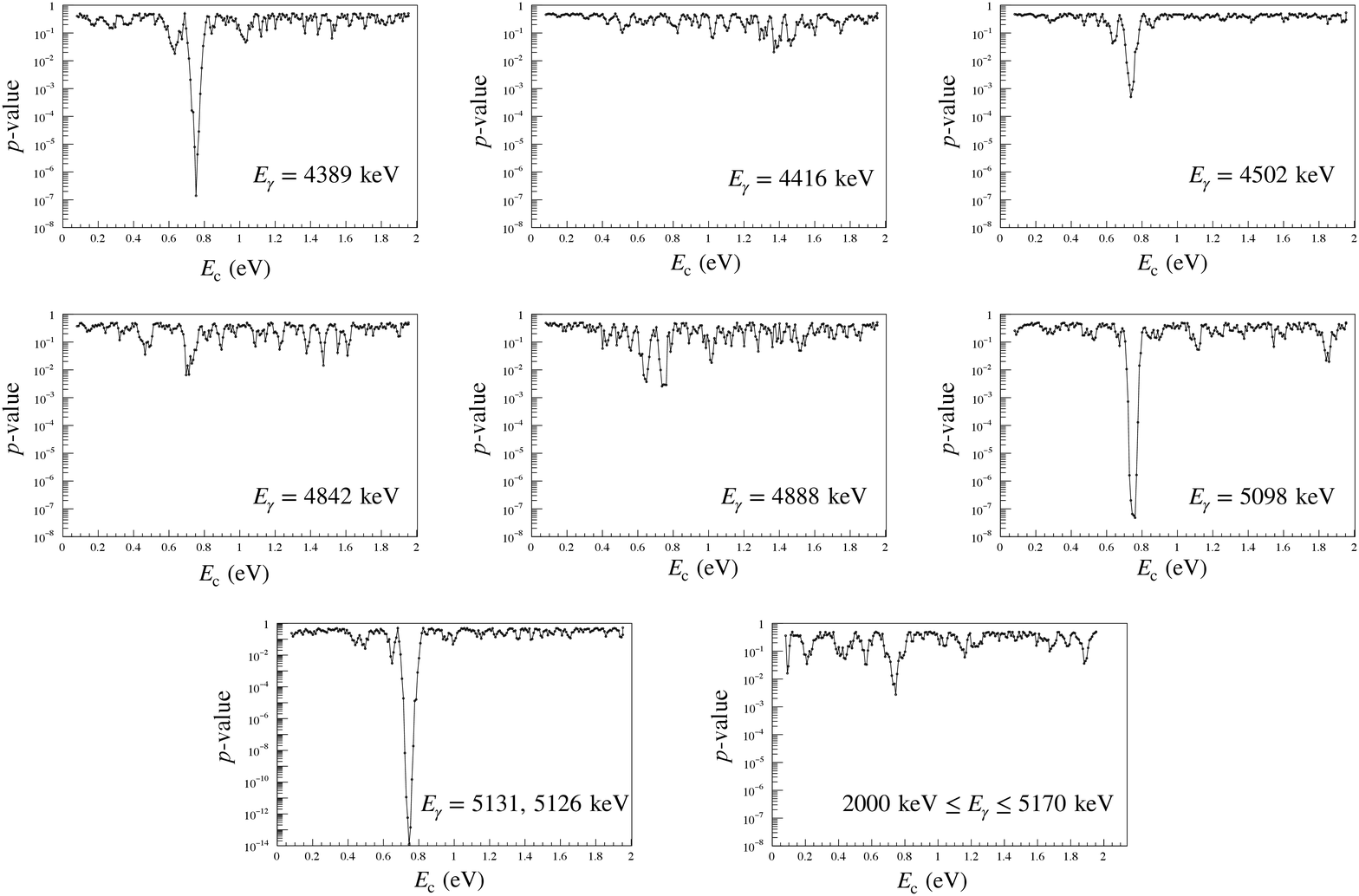}
	\caption[]{{\it p}-values of the angular distribution of $A_{\rm{LH}}$ as a function of the center energy of the integral. }
	\label{pvalue}
	\end{center}
\end{figure*}
\subsection{Significance of the angular distribution}
In order to confirm that the angular distributions are observed at the p-wave resonance, the angular distributions of $A_{\rm{LH}}$ were obtained for other neutron energies, not only within the vicinity of the p-wave resonance. The asymmetry $A_{\rm{LH}}$ is calculated using $N_{\rm L}$ and $N_{\rm H}$ with the integral regions of $E_{\rm{c}}-2\Gamma_{\rm 2} \le E_{n} \le E_{\rm{c}}$ and $E_{\rm{c}}\le E_{n} \le E_{\rm{c}}+2\Gamma_{\rm 2}$, respectively, where the center energy of the integral $E_{\rm{c}}$ takes a value of every 20~meV from 0~eV to 2~eV, and then the angular dependence $A$ was obtained for every integral region. In Fig.~\ref{pvalue}, the significance of $A$ is measured by a {\it{p}}-value, defined as $p=(1-\rm{C.L.})/2$, where C.L. is the confidence level of the non-zero angular dependence. The {\it p}-value indicates the probability to observe a non-zero value of $A$ in the hypothesis of no angular distributions. A confidence level of over 99.7\% corresponds to a {\it p}-value less than $1.35 \times 10^{-3}$. As shown in the graph for $E_\gamma$=4389~keV,  4502~keV, 5098~keV, and 5126~keV and/or 5131~keV in Fig.~\ref{pvalue}, significant angular distributions are observed only within the vicinity of the p-wave resonance. 

This analysis suggests that the angular distribution measurement in the ($n$,$\gamma$) reaction is sensitive to search for faint $\gamma$-ray transitions from p-wave resonances.

\section{Conclusion}
We observed significant angular distributions depending on the neutron energy for the $\gamma$-rays in the transitions from the p-wave resonance of $^{139}$La+$n$ to the several lower excited states of $^{140}$La, including faint $\gamma$-ray transition from the p-wave resonance. This angular distribution can be interpreted as a result of the interference between s- and p-wave amplitudes. Recently, a transverse asymmetry has been measured for the p-wave resonance in the $^{139}$La($n$, $\gamma$)$^{140}$La reaction by using polarized epi-thermal neutrons~\cite{Yamamoto20}, and these measurement results will be combined in terms of the s-p mixing model in order to understand the reaction mechanism of the enhancement of the symmetry violation to be published in a separate paper.

\begin{acknowledgments}
The authors would like to thank the staff of beamline04 for the maintenance of the germanium detectors, and MLF and J-PARC for operating the accelerators and the neutron production target.  The neutron scattering experiment was approved by the Neutron Scattering Program Advisory Committee of IMSS and KEK (Proposal Nos. 2014S03, 2015S12). The neutron experiment at the Materials and Life Science Experimental Facility of the J-PARC was performed under a user program (Proposal Nos. 2016B0200, 2016B0202, 2017A0158, 2017A0170, 2017A0203). This work was supported by MEXT KAKENHI Grant No. JP19GS0210, JSPS KAKENHI Grant Nos. JP17H02889, JP19K21047, and 20K14495.
\end{acknowledgments}
\bibliography{ngamma}
\end{document}